\documentclass[a4paper]{article}

\usepackage{INTERSPEECH2019}

\title{DigiVoice: Voice Biomarker Featurization and Analysis Pipeline }
\name{Larry Zhang$^1$, Xiaotong Chen$^1$, Abbad Vakil$^1$, Ali Byott$^1$, Reza Hosseini Ghomi$^1$}
\address{
  $^1$DigiPsych Lab, Department of Neurology, University of Washington, Seattle, WA, USA
  }
\email{lazhang@uw.edu, xichen1@uw.edu, abbadv@uw.edu, ,rezahg@uw.edu}

\begin{document}

\maketitle

\begin{abstract}
In recent years, data-driven models have enabled significant advances in medicine. Simultaneously, voice has shown potential for analysis in precision medicine as a biomarker for screening illnesses. There has been a growing trend to pursue voice data to understand neuropsychiatric diseases. In this paper, we present DigiVoice, a comprehensive feature extraction and analysis pipeline for voice data. DigiVoice supports raw .WAV files and text transcriptions in order to analyze the entire content of voice. DigiVoice supports feature extraction including acoustic, natural language, linguistic complexity, and semantic coherence features. DigiVoice also supports machine learning capabilities including data visualization, feature selection, feature transformation, and modeling. To our knowledge, DigiVoice provides the most comprehensive voice feature set for data analysis to date. With DigiVoice, we plan to accelerate research to correlate voice biomarkers with illness to enable data-driven treatment. We have worked closely with our industry partner, NeuroLex Laboratories, to make voice computing open source and accessible. DigiVoice enables researchers to leverage our technology across the domains of voice computing and precision medicine without domain-specific expertise. Our work allows any researchers to use voice as a biomarker in their past, current, or future studies.
\end{abstract}


\section{Introduction}

The use of voice data and voice computing technologies has been around for decades. More recently, there have been more studies published discussing the use of voice features to predict illness, including the common cold \cite{c1}, depression and affective states \cite{c2}, and Parkinson’s Disease \cite{c3}. The use of voice computing for medical applications is appealing for several reasons. Voice offers a biomarker which is inherently non-invasive, low-cost, accessible, easy, and fast. Specifically, voice can be collected remotely, reducing many barriers to existing screening tests for patients, many of whom need to travel large distances at significant cost to have access to testing. It is lower cost than most screening tests which require access to complex technology (MRI which may cost \$2,000-5,000 out of pocket), or a physician rating scale requiring an in-person and time intensive interview (e.g. Unified Parkinson’s Disease Rating Scale or Hamilton Depression Rating Scale). Given the reduction of these barriers, voice has the potential to reach under-served minority populations, typically marginalized by modern healthcare research. 

In this paper, we introduce the DigiVoice voice pipeline, an open source pipeline to analyze voice data and build digital biomarkers of disease. The DigiVoice pipeline centralizes multiple open-source methods to produce feature sets, and is able to be executed on any computing platform. The DigiVoice pipeline is built upon API's (Application Programming Interface) to encapsulate the design of functions to enable featurization and analysis capabilties.Through the DigiVoice pipeline, we seek to simplify the process of extracting voice features and modeling of extracted data to produce transparent, comprehensive results which can then be used to gain insights into the vocal phenotype of disease to then build predictive algorithms for the disease of interest. The DigiVoice pipeline is a fundamental step into accelerating the research and discoveries that have been made possible through digital vocal biomarkers. 

Notably, the first text describing the field of Voice Computing as it pertains to building biomarkers with voice was written by Jim Schwoebel, CEO of NeuroLex Laboratories, and is titled “Introduction to Voice Computing in Python”. This text informed our work and also has an open-source repository available online(https://github.com/jim-schwoebel/voicebook). The DigiVoice Voice Pipeline can be found at the following github link (https://github.com/NeurolexDiagnosticsVoice-Analysis-Pipeline) , which can be used out-of-box. The DigiVoice Voice Pipeline is built in Python 3 , and can be used across the predominant operating systems (OSX, Linux, Windows).

\section{Feature Extraction Pipeline}

The DigiVoice Feature Extraction Pipeline extracts acoustic, natural language, linguistic complexity, and semantic coherence features. The features implemented in the pipeline have been chosen as a result of comprehensive literature reviews where these features were highlighted as predictive variables for multiple illnesses, particularly within neuropsychiatric illnesses. Our work discusses our features in more detail as explored pertaining to each specific disease process.

\subsection{Acoustic Features}

The acoustic feature extraction library enables the capability of extracting GeMAPS (Geneva Minimal Acoustic Parameter Set), AVEC2013 (Audio-Visual Emotion recognition Challenge), and Librosa Spectral features from raw 16-bit PCM(pulse-code modulation) .WAV files. These acoustic features capture key signals of voice including Pitch (fundamental (F0) Frequency), Jitter, Shimmer, and Mel-Frequency Cepstral Coefficients(MFCC). 

\subsubsection{Gemaps Features}
The GeMAPS Feature API leverages the OpenSmile feature extraction toolkit to obtain the corresponding acoustic features\cite{c4}. In our pipeline, we leverage the eGeMAPS Parameter Set (extended GeMAPS), consisting of 90 features. The Gemaps parameter set is discussed in depth in Eyben et al . GeMAPS features have been used previously in several studies to capture the predictive power of acoustic features in predicting illnesses and health outcomes. 

In Wroge et al., GeMAPS features were leveraged to classify the presence of Parkinson's Disease via supervised machine learning algorithms\cite{c5}. The paper demonstrated a peak accuracy of 85\% in classifying presence of Parkinson's Disease, which exceeds the average clinical diagnosis accuracy of 73.8\% . In Xing et al., GeMAPS features were leveraged to predict the presence of Bipolar Disorder using Gradient Boosted Decision Tree Classifiers with 69.84\% Accuracy\cite{c6}. Based on these studies, acoustic features may be valuable in detecting functional changes in the central nervous system affecting psycho-motor function (e.g. voice).

\subsubsection{Avec2013 Features}
The AVEC2013 Feature API leverages the OpenSmile feature extraction toolkit to obtain the corresponding acoustic features. Just as GeMAPS Features can be used to detect psycho-motor function, AVEC features can be substituted to accomplish the same goal. In Williamson et al., researchers used a subset of AVEC features to predict changes in dynamic motor function in depressed patients, with a RMSE of 7.42 \cite{c7}. The AVEC2013 parameter contains 2268 features and consists of various low level descriptors (e.g. summary statistics, minima and maxima values, regression functionals) on top of the base set of features (e.g. F0,Jitter,Shimmer,HNR,MFCC,Zero Crossing Rate,Energy,Entropy). A more formal summary of features can be found in Valstar et. Al\cite{c2}. 

\subsubsection{Librosa Features}
The Librosa Spectral Features leverages the Librosa Python Package \cite{c8}. This package is able to capture key information about mel-frequency cepstral coefficients (MFCCs) and spectrogram features. Although both the GeMAPS and AVEC2013 feature extractions contain spectral features and MFCCs in their parameter set, Librosa provides a more specific parameter set with fewer features which have been used in other analyses to classify and identify neuropsychiatric illnesses.

In Cummins et al., MFCCs were used to classify depression in patients, with an accuracy of 80\% \cite{c9}. The second MFCC, specifically, was identified in a study of depressed patients to be indicative of depression with a classification accuracy of 81.2\% \cite{c10}. 
%%%%%%%%%%%%%%%%%%%%%%%%%%%%%%%%%Librosa Table%%%%%%%%%%%%%%%%%%%%%%%%%%%%%%%%%%
\begin{table}[t]
  \caption{Librosa Features}
  \label{tab:word_styles}
  \centering
  \begin{tabular}{ll}
    \toprule
    \textbf{Feature Type}      & \textbf{Specific Features}                \\
    \midrule
    Spectral Features        & Centroid, Bandwidth; Contrast; Flatness;    \\
                             & Roll-off Frequency; Spectral Flux Onset;    \\
                             & RMSE of Spectrogram Mean, Std.Dev.          \\
                             &,Min, Max, Median;Zero Crossing Rate;        \\
                             & Polynomial Features                         \\\\
    Rhythm Features          & Tempo; Tempogram                            \\
    \bottomrule
  \end{tabular}
\end{table}
%%%%%%%%%%%%%%%%%%%%%%%%%%%%%%%%%%%%%%%%%%%%%%%%%%%%%%%%%%%%%%%%%%%%%%%%%%%%%%%%
%%%%%%%%%%%%%%%%%%%%%%%%%%%%%Natural Language Table%%%%%%%%%%%%%%%%%%%%%%%%%%%%%
\begin{table}[t]
  \caption{Natural Language Features}
  \label{tab:word_styles}
  \centering
  \begin{tabular}{ll}
    \toprule
    \textbf{Feature Type}      & \textbf{Specific Features}                \\
    \midrule
    Part of Speech       & \textbf{Adjective} (Comparative,Superlative); \\
                         & \textbf{Noun} (Singular,Mass,Plural, Proper); \\
                         & \textbf{Conjunction} (Preposition, Subordinate\\
                         & Coordinating);                       \\
                         & \textbf{Pronoun} (Personal,Possessive);       \\
                         & \textbf{Adverb} (Comparative,Superlative );   \\
                         & \textbf{Verb}(Base,Past Tense,Past Particle,  \\
                         & Gerund, 3rd person singular);        \\
                         & Determiner, Existential There; Modal;\\
                         & Predeterminer; List Item Marker;     \\
                         & Foreign Word; Cardinal Number; Symbol\\
                         & Interjection; Numeral; Auxiliary     \\\\
    Dependency           & \textbf{Modifiers} (Clausal,Adverbial,Noun  \\
                         & Adjectival,Appositional,Meta,Negation \\
                         & Noun Compound,Possesion,Prepositional,\\
                         & Relative Clause);                    \\
                         & Agent; Attribute; Case Marking; Compound;\\
                         & Conjunct; Copula; Clausal Subject;\\
                         & Dative; Unclassified Dependent; \\
                         & Determiner; Direct Object; Expletive \\
                         & Interjection; Marker; Nominal Subject;\\
                         & Object Predicate; Object; Parataxis; \\
                         & Oblique Nominal;Preposition complement;\\
                         & Object of Preposition; Punctuation;\\
                         & Particle; Root; Open Clausal Complement\\\\
    Other                & Sentiment Value                     \\
                        
    \bottomrule
  \end{tabular}
\end{table}
%%%%%%%%%%%%%%%%%%%%%%%%%%%%%%%%%%%%%%%%%%%%%%%%%%%%%%%%%%%%%%%%%%%%%%%%%%%%%%%%
\subsection{Natural Language Features}
Natural language features provide measurable metrics of lexical composition , which have been shown to have high predictive value in detecting changes in psycho-motor function. Psycho-motor function can provide insight into deterioration in brain function, which could be caused by illnesses including Parkinson's Disease and Alzheimer's Disease. In Orimaye et al., linguistic features were used to predict dementia outcomes via verbal utterances\cite{c11}. Using Support Vector Machine models, they were able to achieve an F-score of 74.9\%. Verbal dependency, sentence structures, and word values were among the features used to predict presence of dementia in contributing samples in the DementiaBank data\cite{c12}. Similar features were used in assessing cognitive function in Parkinson's Disease using the spaCy syntactic dependency parser to obtain part-of-speech tags and dependency features\cite{c13}.

The natural language extraction library uses key natural language packages in python such as nltk and spaCy. The features extracted from nltk include part-of-speech tags, sentiment values, sentence structures, and parts-of-speech dependency factors. In order to extract these features from voice, users will need to have transcriptions of the audio files. Users can use several highly accurate available services to provide manual and automatic transcriptions.

\subsection{Linguistic Complexity Features}
Linguistic Complexity Features are geared towards detecting problems with the flow of conversation and measuring how well subjects can answer questions, or carry conversations without confusion. The linguistic complexity features includes unintelligible word ratio, standardized word entropy, suffix ratio, number ratio, Brunet’s index, Honore’s statistic, and Type-Token ratio. The linguistic complexity features are identified and derived from instructions from Khodabakhsh et al.\cite{c14}. In their paper, they were able to demonstrate a correlation between linguistic complexity and the presence of Alzheimer's Disease. 

\begin{itemize}
\item \textbf{Unintelligible Word Ratio:} The unintelligible word ratio represents the ratio between unintelligible words to intelligible words being spoken in an audio transcript.
\item \textbf{Standarized Word Entropy:} Standardize word entropy is a measure of the word variety and word combination variety that is useful to measure language capacity. The degradation of words and word combinations a patient uses can be indicative of damage in the brain.
\item \textbf{Suffix Ratio:} The suffix ratio gives a measure of the number of suffixes as opposed to other part of speech values. High frequency usage of suffixes reflects a more advanced vocabulary.
\item \textbf{Number Ratio:} The number ratio provides a measure of how many numbers an individual says compared to other words. 
\item \textbf{Brunet's Index:} Brunet's index quantifies lexical richness. A lower Brunet's Index value corresponds to richer lexical content. Brunet's index can be used to gauge a subjects capacity for carrying rich conversations.
\item \textbf{Honore's Statistic:} Honore's Statistic quantifies an individuals lexicon via the notion of individual usage of words. Higher quantities of individual usage of words correlates to richer overall lexicon, and as a result a higher Honore's Statistic
\item \textbf{Type-Token Ratio:} Type-Token ratio defines the ratio of unique words to the total number of words. Similar to Honore's Statistic, Type-Token Ratio measures lexical richness by the uniqueness of words expressed.
\end{itemize}

\subsection{Semantic Coherence Features}
The Semantic Coherence Feature Library is derived from a study conducted by researchers in Bedi et al.\cite{c15}. Inspired by Elvevag et al.\cite{c16}, Bedi and collaborators utilized Latent Semantic Analysis(LSA) to represent the conceptual content of text transcripts. LSA provides a tool to test cognitive function to determine coherence of thought and expression of ideas. They demonstrated a sensitive and specific convex hull model classifier of psychosis (100\% accuracy) using a LSA feature (semantic coherence), maximum phrase length, and rate of determiner use. One of the authors, Carillo, has provided an open-source implementation of semantic coherence (https://github.com/facuzeta/coherence). We have implemented the semantic coherence feature in our pipeline.

%%%%%%Semantic Coherence%%%%%%%%%%
\begin{table}[t]
  \caption{Semantic Coherence Features}
  \label{tab:word_styles}
  \centering
  \begin{tabular}{ll}
    \toprule
    \textbf{Feature Type}      & \textbf{Summary Statistics}                \\
    \midrule
    Semantic Coherence Values  & Mean; Std. Dev.; Minimum;\\
    (0\textsuperscript{th}1\textsuperscript{st}2\textsuperscript{nd}3\textsuperscript{rd}) Order  & Maximum; 10\textsuperscript{th} Percentiles; \\
    & Normalized Summary Statistics\\
    \\ 
    Syntactic Markers & Maximum Phrase Length;\\
    & Use of Determiners\\
    \bottomrule
  \end{tabular}
\end{table}
%%%%%%%%%%%%%%%%%%%%%%%

\section{Data Science Pipeline}
In addition to the feature extraction libraries we are providing through DigiVoice, we have also included a Data Science Pipeline, to enable quick modeling of collected data. The DigiVoice Data Science Pipeline provides data visualization, feature transformation, and feature selection methods. Users will be able to build quick baseline models with their data. The Data Science Pipeline is built upon sci-kit learn methods \cite{c17} and creates a user friendly interface where users can interpret their results.  

\subsection{Data Visualization}
To give a direct and intuitive overview of the data, the data science pipeline supports data visualization methods, including scatterplot, scatterplot matrix, swarm plot, and correlation heat map.
\begin{itemize}

\item \textbf{Scatterplot:} The pipeline provides a scatterplot visualization with corresponding marginal histograms. The plot adds regression and kernel density fits, which are useful to examine the linear relationship between two features
\item \textbf{Scatterplot Matrix:} The scatterplot matrix shows the relationship between every pair of features. The method is similar to the standard scatterplot, however can be used across multiple features
\item \textbf{Swarm Plot:} The swarm plot visualization outputs a categorical scatterplot, which groups swarm plots of each feature. Each color in the swarm plot matches with a categorical target class, which can be useful to observe distribution differences across categories.
\item \textbf{Correlation Heat Map:} The correlation heat map provides a visualization of the correlation matrix of the data. The color gradient varies based upon the strength of the correlation between variables. It can be used to examine potential features with strong correlations amongst themselves \cite{c18}.

\end{itemize}

\subsection{Data Transformation}
The Data Science API also provides a set of data transformation functions, which serve to provide dimensionality reduction methods and addresses collinearities in the data\cite{c19}. Data transformation enables rapid modeling of data due to dimensionality reduction, but reduces the interpretability of the data. It is important to consider the modeling trade-offs when using data transformation methods.

\begin{itemize}

\item \textbf{Low Variance Filter:} The low variance filter method removes features with no variance as they do not contribute to the variance represented in the target class.
\item \textbf{High Correlation Filter:} The high correlation filter method removes groups of features with high correlation to each other to reduce redundancy. The filter retains one feature in the group in case the feature is relevant to the target class.
\item \textbf{Principal Component Analysis (PCA):} The principal component analysis method extracts a set of principal components from existing features. The function accepts an input value of the number of principal components desired. Users can control the dimensions of the transformed data with this method \cite{c20}. 
\item \textbf{Independent Component Analysis (ICA):} The independent component analysis is similar to PCA. ICA attempts to obtain independent features as opposed to uncorrelated features with PCA\cite{c21}.
\item \textbf{Factor Analysis:} The factor analysis method groups highly correlated together as a set of factors. The factor analysis reveals latent variables with covariance\cite{c22}.

\end{itemize}

\subsection{Feature Selection}
The Data Science API provides a set of feature selection functions that users can use to reduce the number of features. Feature selection methods limit the total number of features that are used to model data. Feature selection methods maintain interpretability of data, as opposed to feature transformation methods.

The data science pipeline supports several feature selection methods including Univariate Feature Selection, Recursive Feature Elimination (RFE), Selection via Feature Importance, and the Minimum Redundancy Maximum Relevancy method.

The data post-processed by the data science pipeline is then fed into regression/ classification models depending on problem type. The pipeline provides a plot based on the cross-validation scores provided by the final model against the number of features selected. 

\begin{itemize}

\item \textbf{Univariate Feature Selection:} The univariate feature selection method accepts the number of features to select and provides the output vector with the selected feature sets based upon ANOVA F-values. Univariate feature selection can only be used in classification. Univariate feature selection is a common filter method for feature selection and is relatively efficient \cite{c23}.
\item \textbf{Recursive Feature Elimination (RFE):} The recursive feature elimination method accepts a number of features to select and provides output vector via cross-validated selection based on a greedy search to identify best feature subsets. Recursive feature elimination is computationally intensive, and scales exponentially with respect to the size of the data. To reduce the run-time of the function, our pipeline changes steps based on the size of the data. The accuracy is affected, but run-time is reduced \cite{c24}.
\item \textbf{Selection via Feature Importances:} This method uses the SelectFromModel method from python's scikit-learn package. It leverages baseline models (including LASSO regression) to obtain feature importance. The data transformer will output a set of reduced data based on a threshold upon the feature importance coefficients.
\item \textbf{Minimum Redundancy Maximum Relevancy (MRMR):} The MRMR function ranks features based on maximum relevance with respect to target classes, and minimizes redundancy with respect to other features. The relevance of a feature is calculated as the correlation between itself and the target class. The redundancy of a feature is calculated as the mean absolute value of correlations between the feature and other features. The usage of correlations in the MRMR function only focuses on the relevance and redundancy of linear relationships in the data \cite{c25}.

\end{itemize}

\section{Conclusions}

The DigiVoice Pipeline provides an easy to use, comprehensive approach for leveraging voice computing for research. In the future, we plan to add other key speech features such as speaker diarization, voice activity detection, prosodic features, and automated transcription to further enable end-to-end seemless usage. Through making the DigiVoice pipeline openly available, we hope to enable collaborative, democratized usage of voice technology in health-related analyses.

\section{Acknowledgements}

We would like to acknowledge NeuroLex Laboratories in their assistance in helping us develop the pipeline and making voice computing openly accessible. We would also like to acknowledge the audEERING team for providing the open-sourced openSMILE toolkit. Dr. Hosseini Ghomi’s work was supported by the VA Advanced Fellowship Program in Parkinson’s Disease and NIH R25 MH104159. This funding provided protected time in Dr. Hosseini Ghomi’s schedule for research.

\section{Disclosures}
Dr. Hosseini Ghomi is a stockholder of NeuroLex Laboratories.

\bibliographystyle{IEEEtran}

\begin{thebibliography}{9}
\bibitem{c1} M. Huckvale and A. Beke, “It Sounds Like You Have a Cold! Testing Voice Features for the Interspeech 2017 Computational Paralinguistics Cold Challenge,” in Interspeech 2017, 2017, pp. 3447–3451.
\bibitem{c2} M. Valstar et al., “AVEC 2013: the continuous audio/visual emotion and depression recognition challenge,” in Proceedings of the 3rd ACM international workshop on Audio/visual emotion challenge - AVEC ’13, Barcelona, Spain, 2013, pp. 3–10.
\bibitem{c3} L. Moro-Velázquez, J. A. Gómez-García, J. I. Godino-Llorente, J. Villalba, J. R. Orozco-Arroyave, and N. Dehak, “Analysis of speaker recognition methodologies and the influence of kinetic changes to automatically detect Parkinson’s Disease,” Applied Soft Computing, vol. 62, pp. 649–666, Jan. 2018.
\bibitem{c4} F. Eyben et al., “The Geneva Minimalistic Acoustic Parameter Set (GeMAPS) for Voice Research and Affective Computing,” IEEE Transactions on Affective Computing, vol. 7, no. 2, pp. 190–202, Apr. 2016.
\bibitem{c5} T. J. Wroge, Y. Ozkanca, C. Demiroglu, D. Si, D. C. Atkins, and R. H. Ghomi, “Parkinson’s Disease Diagnosis Using Machine Learning and Voice,” in 2018 IEEE Signal Processing in Medicine and Biology Symposium (SPMB), Philadelphia, PA, 2018, pp. 1–7.
\bibitem{c6} X. Xing, B. Cai, Y. Zhao, S. Li, Z. He, and W. Fan, “Multi-modality Hierarchical Recall based on GBDTs for Bipolar Disorder Classification,” in Proceedings of the 2018 on Audio/Visual Emotion Challenge and Workshop  - AVEC’18, Seoul, Republic of Korea, 2018, pp. 31–37.
\bibitem{c7} J. R. Williamson, T. F. Quatieri, B. S. Helfer, R. Horwitz, B. Yu, and D. D. Mehta, “Vocal biomarkers of depression based on motor incoordination,” in Proceedings of the 3rd ACM international workshop on Audio/visual emotion challenge - AVEC ’13, Barcelona, Spain, 2013, pp. 41–48.

\bibitem{c8} B. McFee et al., “librosa: Audio and Music Signal Analysis in Python,” presented at the Python in Science Conference, Austin, Texas, 2015, pp. 18–24.
\bibitem{c9} N. Cummins, “An investigation of depressed speech detection: features and normalization,” in in Interspeech, 2011, pp. 6–9.
\bibitem{c10} T. Taguchi et al., “Major depressive disorder discrimination using vocal acoustic features,” Journal of Affective Disorders, vol. 225, pp. 214–220, Jan. 2018.
\bibitem{c11} S. O. Orimaye, J. S.-M. Wong, and K. J. Golden, “Learning Predictive Linguistic Features for Alzheimer’s Disease and related Dementias using Verbal Utterances,” in Proceedings of the Workshop on Computational Linguistics and Clinical Psychology: From Linguistic Signal to Clinical Reality, Baltimore, Maryland, USA, 2014, pp. 78–87.
\bibitem{c12} J. T. Becker, F. Boller, O. L. Lopez, J. Saxton, and K. L. McGonigle, “The natural history of Alzheimer’s disease. Description of study cohort and accuracy of diagnosis,” Arch. Neurol., vol. 51, no. 6, pp. 585–594, Jun. 1994.

\bibitem{c13} L. Jessiman, G. Murray, and M. Braley, “Language-Based Automatic Assessment of Cognitive and Communicative Functions Related to Parkinson’s Disease,” in Proceedings of the First International Workshop on Language Cognition and Computational Models, Santa Fe, New Mexico, USA, 2018, pp. 63–74.
\bibitem{c14} A. Khodabakhsh, F. Yesil, E. Guner, and C. Demiroglu, “Evaluation of linguistic and prosodic features for detection of Alzheimer’s disease in Turkish conversational speech,” J AUDIO SPEECH MUSIC PROC., vol. 2015, no. 1, p. 9, Mar. 2015.
\bibitem{c15} G. Bedi et al., “Automated analysis of free speech predicts psychosis onset in high-risk youths,” npj Schizophrenia, vol. 1, no. 1, Dec. 2015.
\bibitem{c16} B. Elvevåg, P. W. Foltz, M. Rosenstein, and L. E. DeLisi, “An automated method to analyze language use in patients with schizophrenia and their first-degree relatives,” J Neurolinguistics, vol. 23, no. 3, pp. 270–284, May 2010.
\bibitem{c17} F. Pedregosa et al., “Scikit-learn: Machine Learning in Python,” MACHINE LEARNING IN PYTHON, p. 6.
\bibitem{c18} W. F. Kuhfeld, “Heat Maps: Graphically Displaying Big Data and Small Tables,” p. 20.
\bibitem{c19} P. Sharma, “Comprehensive Guide to 12 Dimensionality Reduction Techniques,” Analytics Vidhya, 26-Aug-2018. 
\bibitem{c20} J. Shlens, “A Tutorial on Principal Component Analysis,” p. 13.
\bibitem{c21} A. Hyvärinen and E. Oja, “Independent component analysis: algorithms and applications,” Neural Networks, vol. 13, no. 4–5, pp. 411–430, Jun. 2000.
\bibitem{c22} A. B. Costello and J. W. Osborne, “Best Practices in Exploratory Factor Analysis: Four Recommendations for Getting the Most From Your Analysis,” Exploratory Factor Analysis, vol. 10, no. 7, p. 9.
\bibitem{c23} T. Khoshgoftaar, D. Dittman, R. Wald, and A. Fazelpour, “First Order Statistics Based Feature Selection: A Diverse and Powerful Family of Feature Seleciton Techniques,” in 2012 11th International Conference on Machine Learning and Applications, 2012, vol. 2, pp. 151–157.
\bibitem{c24} X. Chen and J. C. Jeong, “Enhanced recursive feature elimination,” in Sixth International Conference on Machine Learning and Applications (ICMLA 2007), 2007, pp. 429–435.
\bibitem{c25} C. Ding and H. Peng, “Minimum Redundancy Feature Selection from Microarray Gene Expression Data,” p. 8.
\end{thebibliography}

\end{document}